\documentclass[aps,prb,preprint,groupedaddress]{revtex4}

\usepackage{graphicx}% Include figure files
\usepackage{dcolumn}% Align table columns on decimal point
\usepackage{bm}% bold math
\usepackage[nice]{nicefrac}
\begin{document}
%{\bf Submitted to Physical Review B, 2005}

% Use the \preprint command to place your local institutional report
% number in the upper righthand corner of the title page in preprint mode.
% Multiple \preprint commands are allowed.
% Use the 'preprintnumbers' class option to override journal defaults
% to display numbers if necessary
%\preprint{}

%Title of paper
\title{Nonlocal Landau theory of the magnetic phase diagram of highly frustrated magnetoelectric CuFeO$_2$}

% repeat the \author .. \affiliation  etc. as needed
% \email, \thanks, \homepage, \altaffiliation all apply to the current
% author. Explanatory text should go in the []'s, actual e-mail
% address or url should go in the {}'s for \email and \homepage.
% Please use the appropriate macro foreach each type of information

% \affiliation command applies to all authors since the last
% \affiliation command. The \affiliation command should follow the
% other information
% \affiliation can be followed by \email, \homepage, \thanks as well.
\author{M.L. Plumer}
%\email[]{Your e-mail address}
%\homepage[]{Your web page}
%\thanks{}
%\altaffiliation{}
\affiliation{Department of Physics and Physical Oceanography, Memorial University, St. John's, Newfoundland, Canada, A1B 3X7}
%\email[]{Your e-mail address}
%\homepage[]{Your web page}
%\thanks{}
%\altaffiliation{}
%\email[]{Your e-mail address}
%\homepage[]{Your web page}
%\thanks{}
%\altaffiliation{}
%\email[]{Your e-mail address}
%\homepage[]{Your web page}
%\thanks{}
%\altaffiliation{}
%Collaboration name if desired (requires use of superscriptaddress
%option in \documentclass). \noaffiliation is required (may also be
%used with the \author command).
%\collaboration can be followed by \email, \homepage, \thanks as well.
%\collaboration{}
%\noaffiliation

\date{\today}

\begin{abstract}
A nonlocal Landau-type free energy functional of the spin density is developed to model the large variety of magnetic states which occur in the magnetic field-temperature phase diagram of magnetoelectric CuFeO$_2$.  Competition among long-range quadratic exchange, biquadratic anti-symmetric exchange, and trigonal anisotropy terms, consistent with the high-temperature rhombohedral R$\bar{3}$m crystal symmetry, are shown to all play important roles in stabilizing the unusual combination of commensurate and incommensurate spin structures in this highly frustrated triangular antiferromagnet.  It is argued that strong magnetoelastic coupling is largely responsible for the nonlocal nature of the free energy. A key feature of the analysis is that an electric polarization is induced by a canting of the non-collinear incommensurate spin structure.  Application of the model to ordered spin states in the triangular antiferromagnets MnBr$_2$ and NaFeO$_2$ is also discussed. 
\end{abstract}

% insert suggested PACS numbers in braces on next line
% insert suggested keywords - APS authors don't need to do this
%\keywords{}

%\maketitle must follow title, authors, abstract, \pacs, and \keywords
\maketitle
\section{Introduction}
% body of paper here - Use proper section commands
% References should be done using the \cite, \ref, and \label commands
%\section{Introduction}
One of the more intriguing features of the magnetic field-temperature ($H-T$) phase diagram of the rhombohedrally stacked triangular antiferromagnetic CuFeO$_2$ is the occurrence of a field-induced non-collinear incommensurate (IC) spin structure  amongst a multitude of commensurate collinear magnetic ordered states. \cite{petrenko00,mitsuda00,kimura06,terada06,nakajima07,seki07}  Usual Hund's rules suggest that the magnetic Fe$^{3+}$ ions in this semiconductor do not exhibit low-order spin-orbit coupling since $L=0$ (with $S=5/2$).  The origin of the observed $c$ axis magnetic anisotropy is not precisely understood but it is generally agreed that it should be very weak. \cite{petrenko05,whangbo06}  The frustration inherent in triangular antiferromagnets is known to give rise to period-3 (P3) elliptically or helically polarized ({\bf S}) spin structures at $H=0$. A field applied along the anisotropy ({\bf c}) direction typically induces a spin flop transition ${\bf S} \perp {\bf c}$ in weak axial antiferromagnets. \cite{plumer88a,plumer88b,plumer89,plumer94} In the case of CuFeO$_2$, spin non-collinearity occurs only at moderate values of $H$, and high-field spin states are linearly polarized with ${\bf S} || {\bf c}$. Key to the understanding of these unusual features is the very high degree of frustration resulting from not only the triangular geometry but also the long-range exchange interactions, magnetoelastic coupling, and anti-symmetric exchange leading to the magnetoelectric effect.  In our previous work (Ref.~\onlinecite{plumer07}, hereafter referred to as I), the observation of the field-induced sequence of period-4 (P4), IC, period-5 (P5) and P3 basal-plane modulated spin structures at $T=0$ was shown to be a consequence of competition among these disparate interactions.  As the temperature is lowered in zero applied field, CuFeO$_2$ exhibits successive magnetic transitions at $T_{N1}\simeq 14K$ to a collinear IC state followed by a discontinuous transition at $T_{N2}\simeq 11K$ to a collinear P4 spin structure.  A similar sequence of transitions has also been reported in the triangular antiferromagnets MnBr$_2$ and NaFeO$_2$.\cite{sato,mcqueen}

In the present work, a representation of these effects is incorporated into a non-local Landau-type free energy functional of the spin density, $F[{\bf s}({\bf r})]$, that is constructed from symmetry arguments, in order to develop a model of the complex $H-T$ phase diagram.\cite{plumer88b}  The formalism is essentially phenomenological but contains the same type of $T=0$ interactions that are considered in usual spin Hamiltonians.  In the case of CuFeO$_2$, these include in-plane exchange interactions up to third neighbor, $J_1$, $J_2$ and $J_3$ \cite{mekata93,ajiro94,fukuda98,ye07} as well as inter-plane exchange $J'$.\cite{petrenko05,terada07,ye07}   These interactions lead to a minimization of the wave vector dependent exchange interaction $J_{\bf Q}$ near a multicritical point where multiple periodicities are close in energy. In addition to various biquadratic exchange interactions (symmetric and anti-symmetric) and usual axial anisotropy, the rhombohedral R$\bar{3}$m crystal symmetry also allows for the existence of an unusual trigonal interaction which was previously used to explain the magnetic structure anomalies in pure Ho.\cite{trigonal} The importance of this term in stabilizing canted structures is explored here. This formalism also leads to umklapp-type terms in the free energy which are non-zero only if the ordering wavevector is equal to a multiple of the crystal reciprocal lattice vector ${\bf G}$, i.e., $n{\bf Q}={\bf G}$ with $n$ having values 3, 4 and 5 in the present case. Within the Landau formalism, these types of terms are then responsible for the stability of the commensurate phases depending on the values of $T$ and $H$. A Landau-type expansion based on molecular field theory applied to MnBr$_2$ is presented in Ref.\onlinecite{sato} which contains features in common with the present model. A more formal group theoretic approach to understanding spin structures in a variety of multiferroic compounds, with a discussion of CuFeO$_2$, is given in Ref. \onlinecite{harris}.  

The nonlocal formalism naturally leads to fourth-order, and higher, contributions (in ${\bf s}({\bf r})$) to the free energy which are dependent on ${\bf Q}$ and illustrates explicitly that terms which are independently invariant (with respect to all symmetries) have independent coefficients.\cite{walker,plumer88b} This contrasts with the usual {\it local} form of the Landau energy which results in all isotropic terms having the same coefficient at a given power of ${\bf s}$.  From a phenomenological point of view, the strong symmetry arguments are sufficient to assign different values to each of the many coefficients.  However, such nonlocal effects can also be understood to have microscopic origins by considering interactions between magnetic and other degrees of freedom, such as magnetoelastic coupling, which gives rise to nonlocal biquadratic exchange terms.  Spin-lattice coupling is known to be strong in CuFeO$_2$, \cite{plumer07,magneto,terada08,wang08} which provides further justification for the present approach.  

\section{Nonlocal Landau Free Energy}

The development here of a nonlocal Landau-type free energy functional of the spin density $F[{\bf s}({\bf r})]$ for CuFeO$_2$  follows the description given in Ref. \onlinecite{plumer88b} for triangular antiferromagnets and uses the symmetry arguments given in I. Although the expansion is carried out to sixth order, a number of simplifications can be made for the purpose of understanding the phase diagram.  It is convenient to write our model for $F$ as the sum of isotropic and anisotropic contributions in the form
\begin{eqnarray}
F=F_2 + F_4 + F_6 + F_z + F_{CP} + F_K - {\bf m} \cdot {\bf H}
\end{eqnarray}
where the first three terms are the isotropic contributions relevant for all magnetic systems
\begin{eqnarray}
F_2 &=& \frac{1}{2V^2} \int d{\bf r} d{\bf r'} A({\bf r} - {\bf r'}) {\bf s}({\bf r}) \cdot {\bf s}({\bf r'}) \\
F_4 &=& \frac{1}{4V^4} \int d{\bf r_1} d{\bf r_2} d{\bf r_3} d{\bf r_4} B({\bf r_1},{\bf r_2},{\bf r_3},{\bf r_4})                    [{\bf s}({\bf r_1}) \cdot {\bf s}({\bf r_2})] [{\bf s}({\bf r_3}) \cdot {\bf s}({\bf r_4})]\\
F_6 &=& \frac{1}{6V^6} \int d{\bf r_1} \cdot \cdot \cdot d{\bf r_6}
C({\bf r_1},\cdot \cdot \cdot,{\bf r_6}) 
[{\bf s}({\bf r_1}) \cdot {\bf s}({\bf r_2})] [{\bf s}({\bf r_3}) \cdot {\bf s}({\bf r_4})] [{\bf s}({\bf r_5}) \cdot {\bf s}({\bf r_6})]~.
\end{eqnarray}
Temperature dependence enters in the usual Landau treatment of the isotropic second-order term, here generalized to be of the form
\begin{eqnarray}
A({\bf r}) = a k_BT \delta({\bf r}) + j^2J({\bf r})
\end{eqnarray}
where $a$ depends on the total angular momentum number $j$ and $J({\bf r})$ is the exchange interaction.  Such a form can be derived from a molecular field treatment of the corresponding Heisenberg hamiltonian.\cite{sato,bak,plumer91}  This mean field treatment also yields {\it local} forms for the higher order isotropic terms, involving constants $B$ and $C$ which also depend on $j$ and are proportional to $k_BT$.   The nonlocal form of spin interactions, such as biquadratic exchange, can arise from a variety of $n$-body interactions\cite{adler} or indirectly from the coupling of ${\bf s}({\bf r})$ to other relevant degrees of freedom such as lattice or electronic (see the $F_{CP}$ term below).  Nonlocal fourth order contributions to the energy of the form given above that arise from magnetoelastic coupling have been derived.\cite{bergman06,plumer91} In the present model, we set $k_B \equiv 1$,  $j \equiv 1$ and simply treat $a$ and the nonlocal fourth- and sixth-order coefficients as phenomenological parameters of undefined microscopic origin but with the support for their existence in CuFeO$_2$ from the strong spin-lattice coupling and a nontrival spin-polarized electronic structure.\cite{galakhov} 

There are a large number of independent anisotropic contributions invariant with respect to the R$\bar{3}$m crystal rhombohedral symmetry group generators $\{S_6|000\}$ and $\{\sigma_{v}|000\}$. \cite{bradley} Only the essential terms are considered here. Axial anisotropy (${\bf \hat z} || {\bf \hat c})$ is included only at second order in $s$:
\begin{eqnarray}
F_z =  \frac{1}{2V^2} \int d{\bf r} d{\bf r'} J_z({\bf r} - {\bf r'}) s_z({\bf r})  s_z({\bf r'}). 
\end{eqnarray}
Note that although this general form includes both single-ion anisotropy $D$, where ${\bf r} = {\bf r'}$, as well as two-site anisotropic exchange, the model used in I which includes only the latter is adopted here. Since the anisotropy is small, such a distinction has little impact on the results presented below.  Higher order terms such as $({\bf s} \cdot {\bf s})s^2_z$ and $s^4_z$ are omitted here for simplicity.

The stability of the field-induced non-collinear phase at zero temperature was shown in I to be a result of including a biquadratic anti-symmetric exchange term of the form  
\begin{eqnarray}
F_{CP} = -\frac{1}{8V^2A_P} \sum_\alpha  \int d{\bf r} d{\bf r'} \left[ C({\bm \tau}) {\hat \tau}^{\alpha} ({\bf s}({\bf r}) \times {\bf s}({\bf r'})) \cdot {\bf \hat z}\right]^2 
\end{eqnarray}
where ${\bm \tau} = {\bf r} - {\bf r'}$ and $\alpha = x,y$. Such a term is invariant with respect to all relevant symmetries and therefore must exist in the energy.  Contributions to it can also arise as a result of coupling between the electric polarization, lattice and spin vectors in the form\cite{kimura06}
\begin{eqnarray}
F_C = \frac{1}{2V^2} \int d{\bf r} d{\bf r'} C({\bm \tau}) [{\bf P}({\bm \tau}) \times {\bm {\hat \tau}}] \cdot [{\bf s}({\bf r}) \times {\bf s}({\bf r'})]_z 
\end{eqnarray}
in combination with the lowest-order contribution to $F$ in ${\bf P}$  given by\cite{katsura07,syrom} $F_P=\frac{1}{2V^2}A_P\int d{\bf r}d{\bf r'}P({\bm \tau})^2$.  Such cross product interactions are known to stabilize helically polarized spin structures.\cite{dzy64}  Further analysis of the dependence of ${\bf P}$ on ${\bf s({\bf r})}$ is given in the following section.

Systems with trigonal symmetry also support an unusual anisotropy term which couples in-plane and out-of-plane components of the spin vector.\cite{trigonal}  For simplicity we use the local, single-site expression of this effect with a coefficient $K$:
\begin{eqnarray}
F_K = \frac{K}{2V}  \int d{\bf r}  s_z({\bf r}) s_y({\bf r})[3s^2_x({\bf r}) - s^2_y({\bf r})]. 
\end{eqnarray}
This interaction favors linear spin configurations which are canted, having both $z$ and basal-plane components.  The combination of $F_{CP}$ and $F_K$ contributions to the free energy thus favor a canted elliptically polarized state which induces an electric polarization ${\bf P}$ in the basal plane.

\section{Wave Vector Representation} 

Positions within the rhombohedral lattice of the Fe$^{3+}$ magnetic ions are described here through an equivalent simple hexagonal unit cell with three triangular layers $A$, $B$, and $C$, using basis vectors ${\bf w}_A = 0$, ${\bf w}_B = \frac{1}{2} a {\bf \hat x} + \frac{1}{3} b {\bf \hat y} + \frac{1}{3}c{\bf \hat z}$,
${\bf w}_C =  \frac{1}{2} a {\bf \hat x} - \frac{1}{3} b {\bf \hat y} - \frac{1}{3}c{\bf \hat z}$, with $a$ and $c$ being the lattice constants, $b=(\sqrt{3}/2)a$ and ${\bf \hat x}\perp {\bf \hat y} \perp {\bf \hat z}$. 
As discussed previously,\cite{plumer88b,sato,plumer91} long-range magnetic order is characterized by the function ${\bm \rho}({\bf r})$ through the spin density written as
\begin{eqnarray}
{\bf s}({\bf r}) = \frac{V}{N} \sum _{\bf R_l} \sum_j {\bm \rho}_j({\bf r}) \delta({\bf r-R})
\end{eqnarray}
where ${\bf R} = {\bf R_l} + {\bf w}_j$, ${\bf R_l} = (n+ {\textstyle{\frac{1}{2}}}m) a {\bf \hat x} + mb{\bf \hat y}$ (with $n$ and $m$ being integers) specifies hexagonal Bravais lattice sites, $j=A,B,C$, and $N=3N_l$ is the number of magnetic ions. 
The variety of spin structures in the magnetic phase diagram of CuFeO$_3$ can be adequately distinguished with ${\bm \rho}$ expressed as a Fourier expansion with a small number of terms in the form
\begin{eqnarray}
{\bm \rho}_j({\bf r}) = {\bf m} + {\bf S}_j e^{i{\bf Q} \cdot {\bf r}} +  {\bf S}_j^*  e^{-i{\bf Q} \cdot {\bf r}}
\end{eqnarray}
where ${\bf m}$ is the uniform magnetization induced by an applied field along the ${\bf \hat z}$ axis, ${\bf S}$ is the spin polarization vector and ${\bf Q}$ is the primary wave vector. 
A full description of some of the ordered states requires additional Fourier components, as discussed below.

 With only near-neighbor coupling $J'$ between planes, a reasonable {\it ansatz} is to assume that the complex spin polarization vectors ${\bf S}_j$ on adjacent layers differ only by a simple phase factor $\phi$ with each layer having an overall phase factor $\gamma$ in the form\cite{plumer91}
\begin{eqnarray}
 {\bf S}_A = {\bf S} e^{i{\gamma}}, ~~ {\bf S}_B = {\bf S}_C = {\bf S} e^{i({\gamma}-\phi)}.
\end{eqnarray}
The primary description of the spin polarization of each state is further characterized by writing\cite{plumer88a}   
\begin{eqnarray}
 {\bf S} = {\bf S}_1 + i {\bf S}_2  
\end{eqnarray}
where ${\bf S}_1$ and ${\bf S}_2$ are real vectors. Previous analysis of Landau-type models to describe the magnetic states of frustrated triangular antiferromagnets\cite{plumer88a,plumer88b,plumer89,plumer94} have demonstrated that sufficient flexibility can be achieved by assuming ${\bf S}_1$ and ${\bf S}_2$ are characterized by writing
\begin{eqnarray}
{\bf S}_1 = S \cos \beta [\sin \theta {\hat {\bm \rho}}_1 + \cos \theta {\bf \hat z}], ~~ {\bf S}_2 = S \sin \beta {\hat {\bm\rho}}_2 
\end{eqnarray}
where 
${\hat {\bm \rho}_1} \perp {\hat {\bm \rho}_2} \perp {\bf \hat z}$, i.e., ${\hat {\bm\rho}_1}$ and ${\hat {\bm \rho}_2}$ are in the hexagonal basal plane.
The $Q-th$ component of the spin density on layer $A$ is then given by
\begin{eqnarray}
 {\bf s}_A({\bf r}) = 2{\bf S}_1 \cos({\bf Q} \cdot {\bf r} + \gamma) -   2{\bf S}_2 \sin({\bf Q} \cdot {\bf r} + \gamma).
\end{eqnarray}
Linearly polarized states are described with ${\bf S}_2 = 0$, proper helically polarized spin configurations are realized by ${\bf S}_1 \perp {\bf S}_2 \perp {\bf Q}$ with $S_1 = S_2$ ($\beta = \pi/4$), and elliptically polarized states have $S_1 \neq S_2 \neq 0$.
It is also convenient to define dimensionless wave vectors in units of the lattice constants
\begin{eqnarray}
 q_x = a Q_x, ~~ q_y = b Q_y,~~  q_z = c Q_z.
\end{eqnarray}
The zero-field $\uparrow \uparrow \downarrow \downarrow$ P4 phase described in Fig. 11 of Ref.~\onlinecite{mekata93} for a single layer is characterized by the above relations with $q_x=\pi$, $q_y=0$, or equivalently $q_x=\pi /2$, $q_y= 3 \pi / 4 $ due to the triangular symmetry, along with $\beta =0$, $\theta = 0$ and $\gamma = \pi /4$ (also see Ref.\onlinecite{sato}). In terms of triangular lattice basis vectors ${\bf a} =a{\bf \hat x}$ and ${\bf b} =\textstyle{\frac{1}{2}}a{\bf \hat x} + b{\bf \hat y}$, the relations are $q_a=q_x$ and $q_b=\textstyle{\frac{1}{2}}q_x+q_y$.  Thus, for example, the above degenerate modulations are also described by $q_a=\pi$, $q_b=\pi /2$, and $q_a=\pi /2$, $q_b= \pi$.
%  The convention for theoretical descriptions of spin structures with rhombohedral symmetry is the use of Cartesian coordinates.\cite{jqrefs}

\subsection{Second Order Isotropic Terms}
The free energy density as a function of ${\bf m}$, ${\bf S}$ and ${\bf Q}$ is developed by using the spin density given by (10) and (11) in expressions (1)-(9).  As shown previously,\cite{plumer88b,plumer91} this analysis naturally leads to umklapp terms arising from the condition for periodic lattices:
\begin{eqnarray} 
\frac{1}{N_l} \sum_{\bf R_l} e^{in{\bf Q} \cdot {\bf R_l}} = \Delta_{n{\bf Q},{\bf G}}
\end{eqnarray}  
where ${\bf G}$ is a hexagonal reciprocal lattice vector.  Second order isotropic contributions reduce to
\begin{eqnarray} 
F_2 =  {\textstyle \frac{1}{2}} A_0 m^2 + A_{\bf Q} S^2
\end{eqnarray}  
where $S^2 = {\bf S} \cdot {\bf S}^*$, $A_{\bf Q} = aT + J_{\bf Q}$ and
\begin{eqnarray} 
J_{\bf Q} =  \frac{1}{N} \sum_{\bf R} J({\bf R}) e^{i{\bf Q} \cdot {\bf R}}.  
\end{eqnarray}
Within the present model, this leads to the following wave vector dependence of the exchange integral $J_{\bf Q} = 2f({\bf q}, \phi)$, with  
\begin{eqnarray} 
f({\bf q}, \phi) =  J_1 f_1({\bf q}) + J_2 f_2({\bf q}) + J_3 f_3({\bf q}) + {\textstyle\frac{1}{3}}J' f'({\bf q})(1 + 2 \cos \phi)  
\end{eqnarray}
where\cite{jqrefs}  
\begin{eqnarray}
f_1 &=& \cos q_x + 2 \cos   {\textstyle\frac{1}{2}}  q_x \cos q_y \nonumber \\
f_2 &=&  \cos 2 q_y + 2 \cos {\textstyle\frac{3}{2}}  q_x \cos q_y \nonumber \\
f_3 &=&  \cos 2 q_x + 2 \cos q_x \cos 2 q_y  \nonumber \\
f' &=&  \cos ({\textstyle\frac{2}{3}}q_x - {\textstyle\frac{1}{3}q_z)} + 2 \cos {\textstyle\frac{1}{2}}q_x \cos {\textstyle(\frac{1}{3}}q_y + {\textstyle\frac{1}{3}}q_z)  
\end{eqnarray}
and $J_1=J(a{\bf \hat x})$,  $J_2=J(2b{\bf \hat y})$, $J_3=J(2a{\bf \hat x})$, $J'=J({\bf w})$. Note that $J_i >0$ represents antiferromagnetic coupling. Using these results, the coefficient of $m^2$ can be expressed as $A_0 = aT + 2f(0,0)$. 

Within this mean field theory, the wave vector characterizing the first ordered phase to appear as the temperature is lowered from the paramagnetic state ($S=0$) is determined by the extrema of $J_{\bf Q}$ (the effects of anisotropic exchange are discussed below).  Results of a numerical algorithm to sketch the $J_2 - J_3$ phase diagram are shown in Fig. 1 (also see Ref. \onlinecite{jqrefs}).  Here, we set $J_1 \equiv 1$ for convenience and consider two values of inter-layer coupling, $J'=0$ and $J'=0.4$, as in I.  In the case of $J'=0$ (broken lines), the usual P3 modulation found in the frustrated triangular antiferromagnet (associated with, for example, $120^\circ$ spin structures) occurs in the upper left part of the diagram.  At more positive values of $J_2$, further frustration stabilizes an IC modulation.  Simple antiferromagentic structures\cite{mekata93} (P2) are found in the lower right region.  Note that due to the triangular symmetry, there are a number of equivalent wave vector descriptions for the same basic structure.  For example, $120^\circ$ spin configurations with $(4 \pi /3,0)$, $(2 \pi /3,  \pi)$ differ only in their chirality.\cite{plumer88b}  Period-2 structures with $(0,\pi), (2 \pi, 0)$ and $(\pi,  \pi /2)$ differ only by a rotation of axes.     With the addition of interlayer exchange $J'=0.4$ (solid lines), there is a nontrivial interplay between $q_z$ and $q_x, q_y$ as can be seen in Eq. (21).  The boundary between P3 and IC phases disappears and there is an additional AF or IC modulation between adjacent planes.  For example, in the lower part of the phase diagram, degenerate modulations $(0,\pi,2 \pi)$ and $(2\pi,0,0)$ both yield simple AF inter-plane structures since ${\bf q} \cdot {\bf w}_B = \pi$ and ${\bf q} \cdot {\bf w}_C = \pi$.   Also indicated on Fig. 1 is the point corresponding to $J_2=0.3$, $J_3=0.3$  (close to the values in I) which are used in the calculation of the $H-T$ phase diagram described below.  

\begin{figure}[h]
\includegraphics[width=1.0\textwidth]{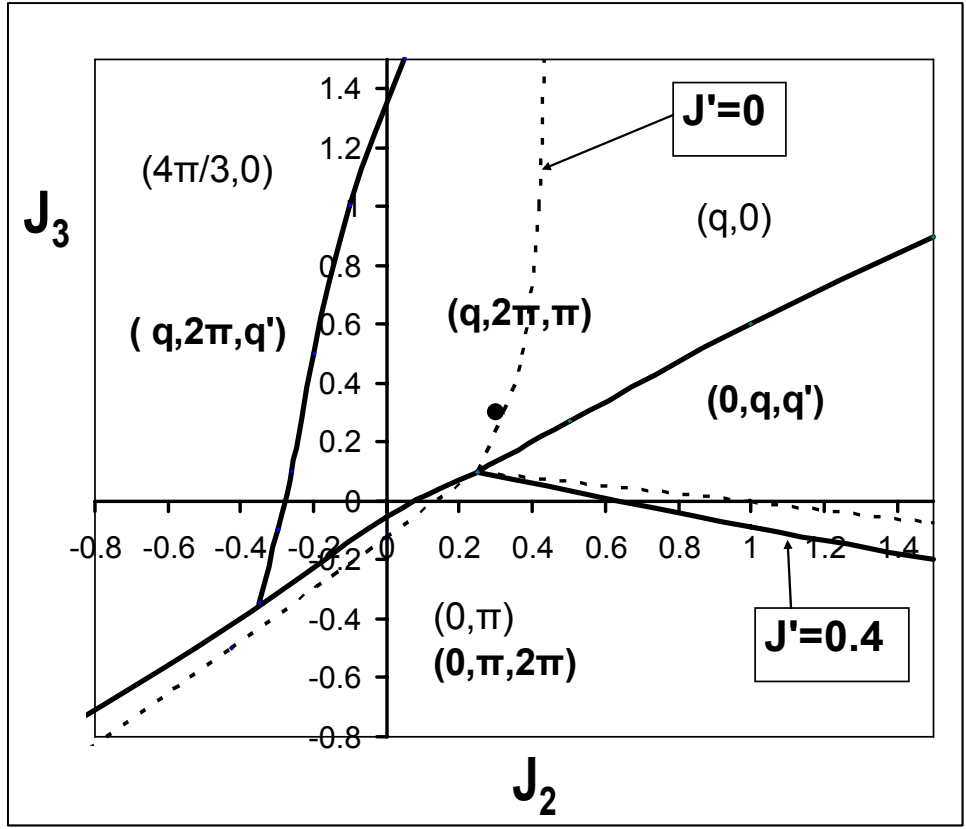}
\caption{\label{one} Sketch of the $J_2 - J_3$ phase diagram based on minimization of the exchange integral $J_{\bf q}$ with $J_1 = 1$.  Broken curves correspond to the case $J'=0$ and solid curves $J'=0.4$.  Solid circle indicates values used in the present model: $J_2 = J_3 = 0.3$, $J'=0.4$.}
\end{figure}

\subsection{Fourth and Sixth Order Isotropic Terms}

Higher order terms can also be evaluated with the assumed spin density (10)-(12).  For convenience, regular (non-umklapp) terms and umklapp terms are written separately.  In the case of the isotropic fourth-order contributions, the result can be expressed as follows:
\begin{eqnarray}
F_4 = F_{4,R} + F_{4,3} + F_{4,4}  
\end{eqnarray}
where
\begin{eqnarray} 
F_{4,R} &=& B_1 S^4 + {\textstyle\frac{1}{2}} B_2 |{\bf S} \cdot {\bf S}|^2 + {\textstyle\frac{1}{4}} B_3 m^4 + 
2B_4 |{\bf m} \cdot {\bf S}|^2 + B_5 m^2 S^2 \\
F_{4,3} &=& B_{4,3} [ ({\bf m} \cdot {\bf S}) ({\bf S} \cdot {\bf S})e^{3i\gamma} + c.c.]\Delta_{3{\bf Q},{\bf G}} \\
F_{4,4} &=& {\textstyle\frac{1}{4}}B_{4,4} [  ({\bf S} \cdot {\bf S})^2 e^{4i\gamma} + c.c.]\Delta_{4{\bf Q},{\bf G}}.
\end{eqnarray}
As shown in Ref.~\onlinecite{plumer88b}, umklapp terms involving $2{\bf Q}$ are accounted for within the regular terms by a suitable renormalization of the spin density amplitude. Expressions for some of the coefficients in terms of the Fourier transform of the nonlocal function $B({\bf r_1},{\bf r_2},{\bf r_3},{\bf r_4})$ are given in Ref. \onlinecite{plumer88b} and the others may be easily deduced.  For example, $B_1 = B_{\bf Q,-Q,Q,-Q}$, $B_4 = B_{\bf 0,Q,0,-Q}$ and $B_{4,3} = B_{\bf 0,Q,Q,Q}$.  The important point here is that the nonlocal formulation naturally leads to the result that each independently invariant term has its own independent coefficient.  In a local formulation, all isotropic terms have equal coefficients.  In the present model, each of these fourth-order (and sixth-order) independent coefficients is assumed to be constant. As discussed previously,\cite{walker,plumer88a} non-collinear spin structures are stabilized with $B_2 > 0$ and collinear states with $B_2<0$.  Usual spin-flop transitions in antiferromagnets occur as a consequence of having $B_4 > 0$ so that a spin configuration with ${\bf S} \perp {\bf H}$ is stabilized. 

The above umklapp terms are non-zero only if $3{\bf Q} ={\bf G}$ (and $m \ne 0$) or $4{\bf Q} ={\bf G}$, allowing for the possibility of energy reduction if the system assumes these periodicities.\cite{sato}  In the case of ${\bf S} || {\bf m}$ for the first term, or in the case of any collinear state for the second term, these expressions reduce to $2B_{4,3}mS^3 \cos(3\gamma)$ and ${\textstyle\frac{1}{2}}B_{4,4} S^4 \cos(4\gamma)$, respectively.  For positive coefficients, these two terms are then each minimized by phase factors $\gamma = \pi / 3$ and $\gamma = \pi / 4$, respectively.  

In an effort to reduce the number of model parameters, a somewhat simplified approach is adopted to treat the isotropic sixth order contributions.  In addition to the regular terms, there are umklapp terms involving periodicities ${\bf G}/3$, ${\bf G}/4$, ${\bf G}/5$ and ${\bf G}/6$ .  For simplicity, it is assumed that each of the independent terms forming the regular part have the same coefficient $C$.  This is equivalent to a local formulation of these contributions. Umklapp terms, however, are each assigned an independent coefficient.  The result can be expressed as 
 
\begin{eqnarray}
F_6 = F_{6,R} + F_{6,3} + F_{6,4} + F_{6,5} + F_{6,6} 
\end{eqnarray}
where
\begin{eqnarray} 
F_{6,R} &=& \textstyle{\frac{1}{6} C  \big\{S_T^6 + 6 S_T^2 |{\bf S} \cdot {\bf S}|^2  + 24 S_T^2 |{\bf m} \cdot {\bf S}|^2 
+ 12[({\bf S} \cdot {\bf S})({\bf m} \cdot {\bf S}^*)^2} + c.c.]\big\} \\
F_{6,3} &=& {\textstyle\frac{1}{6}} C_{6,3}  \bigg\{[12 S_T^2({\bf m} \cdot {\bf S})({\bf S} \cdot {\bf S})
 + 6 ({\bf m} \cdot {\bf S}^*) ({\bf S} \cdot {\bf S})^2 + 8({\bf m} \cdot {\bf S})^3] e^{3i\gamma} + c.c.\bigg\}\Delta_{3{\bf Q},{\bf G}} ~~~~\\
F_{6,4} &=& {\textstyle\frac{1}{6}}C_{6,4} \bigg\{ [3 S_T^2 ({\bf S} \cdot {\bf S})^2 + 20 ({\bf m} \cdot {\bf S})^2 ({\bf S} \cdot {\bf S})] e^{4i\gamma} + c.c.\bigg\} \Delta_{4{\bf Q},{\bf G}}.\\
F_{6,5} &=& C_{6,5} \bigg\{ ({\bf m} \cdot {\bf S}) ({\bf S} \cdot {\bf S})^2e^{5i\gamma} + c.c.\bigg\}\Delta_{5{\bf Q},{\bf G}} \\
F_{6,6} &=& {\textstyle\frac{1}{6}}C_{6,6} \bigg\{  ({\bf S} \cdot {\bf S})^3 e^{6i\gamma} + c.c.\bigg\}\Delta_{6{\bf Q},{\bf G}}.
\end{eqnarray}
Note that odd-order umklapp terms occur only in the presence of a magnetic field where $m \ne 0$ and have a maximum effect on lowering the free energy with linearly polarized spin configurations and ${\bf S} || {\bf H}$.  In-plane and $c$-axis periodicities, n and m, respectively,  can be separated by writing  ${\bf Q} = \textstyle{\frac{1}{n}} {\bf G}_{\perp}$ + $\textstyle{\frac{1}{m}} {\bf G}_{||}$ where ${\bf G}_{\perp}$ and ${\bf G}_{||}$ are reciprocal lattice vectors perpendicular and parallel to the hexagonal $c$ axis.  At low values of magnetic field, antiferromagnetic inter-plane exchange $J'>0$ dominates and period-2 inter-layer $c$ axis modulations are stabilized.
% Both IC and P4 in-plane spin structures are characterized by ${\bf Q}_{||} = \textstyle{\frac{1}{2}}{\bf G}_{||}$ (e.g., $Q_{||} = (3/2) 2\pi /c$), with the umklapp condition ${4{\bf Q} = {\bf G}}$ being satisfied in the latter case.
For the odd-period in-plane commensurate structures, ${\bf Q}_{||} = {\bf 0}, {\bf G}_{||}$ and the umklapp conditions ${3{\bf Q} = {\bf G}}$ and ${5{\bf Q} = {\bf G}}$ are satisfied for P3 and P5 spin configurations, respectively. These characterizations are consistent with neutron diffraction data.\cite{petrenko00,mitsuda00}

A complete description of the spin structures usually requires that additional wave vector components (${\bf Q'}$) be added to the Fourier expansion of the spin density (11) which lead to a `squaring-up' of magnetic structures.\cite{plumer94,sato}  Fourth-order isotropic umklapp terms of the form $({\bf S'}\cdot{\bf S}) ({\bf S} \cdot {\bf S})$ exist for cases where ${\bf Q'} + 3{\bf Q} = {\bf G}$ leading to the incipient relation ${\bf S'}\sim {\bf S} ({\bf S} \cdot {\bf S})$. With a magnetic field present, terms of the form  $({\bf S'}\cdot{\bf m}) ({\bf S} \cdot {\bf S})$ occur at fourth order if ${\bf Q'} + 2{\bf Q} = {\bf G}$ giving  ${\bf S'}\sim {\bf m} ({\bf S} \cdot {\bf S})$.  A larger number of possible secondary wave vectors arise from sixth order umklapp terms.  A fully consistent analysis would lead to many additional contributions to the free energy but are not required for the semi-quantitative description of the phase diagram given here.  

\subsection{Anisotropic Terms}

Evaluating the anisotropic contributions to the free energy,  $F_z$, $F_{CP}$ and $F_K$, in terms of the spin density parameters ${\bf m}$, ${\bf S}$ and ${\bf Q}$ follows the method as described above. Since anisotropy is known to be small in CuFeO$_2$, anisotropic umklapp terms are omitted for simplicity.  The resulting anisotropic exchange terms are identical in form to isotropic exchange, but involve only $z$ components of the spin vectors: 
\begin{eqnarray} 
F_z =  {\textstyle \frac{1}{2}} J_{z0} m_z^2 + J_{z{\bf Q}} |S_z|^2.
\end{eqnarray}  
Here, $J_{z{\bf Q}}$ is given by the relations (19)-(21) but with the isotropic exchange parameters $J_i$ and $J'$ replaced by their anisotropic counterparts $J_{zi}$ and $J_z'$, as in I.  The coefficient $J_{z0}$ is then given by this expression evaluated at ${\bf Q} = {\bf 0}$ and $\phi=0$ in (21). 

 An expression for the nonlocal biquadratic antisymmetric exchange interaction can be determined by first evaluating $F_C$ using (10) and (11) with the assumption that the electric polarization vector ${\bf P}$ is uniform.  This gives\cite{plumer91b}
\begin{eqnarray} 
F_C =   i (C_xP_x + C_yP_y  ){\bf \hat z} \cdot ({\bf S} \times {\bf S}^*)
\end{eqnarray}  
where, as in I, magnetoelectric-type interactions $C_1=C(a{\bf \hat x})$ and $C'=C({\bf w})$ are included, giving
\begin{eqnarray} 
C_x &=& -{\textstyle\frac{4}{3}} b\{  C_1  \cos{\textstyle\frac{1}{2}}q_x \sin q_y - {\textstyle\frac{1}{3}}C' [\sin ({\textstyle\frac{1}{3}}q_z - {\textstyle\frac{2}{3}}q_y)} -  \sin ({\textstyle\frac{1}{3}}q_x + {\textstyle\frac{1}{3}}q_y) \cos {\textstyle{\frac{1}{2}}q_x](1 + 2 \cos \phi) \}  \nonumber \\ 
C_y &=&  {\textstyle\frac{2}{3}} a \{ C_1  (\sin q_x + \sin {\textstyle\frac{1}{2}}q_x \cos q_y)+ C' \sin {\textstyle\frac{1}{2}}q_x \cos ({\textstyle\frac{1}{3}}q_y + {\textstyle\frac{1}{3}}q_z)(1 + 2 \cos \phi) \}.  
\end{eqnarray}  
Minimization of the free energy $F_{CP} = F_C + F_P$, where $F_P= {\textstyle\frac{1}{2}} A_P(P_x^2 + P_y^2)$, thus yields the relations between ${\bf P}$ and the spin polarization vectors:
\begin{eqnarray} 
P_x &=& - (i/A_P) C_x ({\bf S} \times {\bf S}^*)_z \nonumber \\
P_y &=& - (i/A_P) C_y ({\bf S} \times {\bf S}^*)_z.  
\end{eqnarray} 
The combined antisymmetric biquadratic exchange contribution to the free energy then takes the form
\begin{eqnarray} 
F_{CP} =   \frac{1}{2A_p} (C_x^2 + C_y^2)({\bf S} \times {\bf S}^*)_z^2.
\end{eqnarray}  
These relations make clear that a uniform electric polarization cannot be induced by a collinear spin structure where $\beta = 0$ or $\pi /2$ since
\begin{eqnarray} 
({\bf S} \times {\bf S}^*)_z = 2 i ({\bf S}_1 \times {\bf S}_2)_z =   i S^2 \sin 2 \beta \sin \theta.  
\end{eqnarray}  
Note also that ${\bf P} = 0$ in cases where ${\bf S}$ lies strictly in the ${\bf \hat z} - {\hat {\bm \rho}_2}$ plane ($\theta = 0$), as has been recently emphasized.\cite{nakajima07,terada07,arima07,mitamura07}

Finally, the local formulation of the trigonal anisotropy term given above can be expressed in terms of the spin polarization vector components as
\begin{eqnarray}
F_K = K\{[3(S_x^*)^2S_yS_z - S_zS_y(S_y^*)^2 + 2S_yS_z^*(3|S_x|^2 - |S_y|^2)] + c.c.\}. 
\end{eqnarray}
Using the parameterizations of ${\bf S}$ given above, this interaction term can be written as
\begin{eqnarray}
F_K = 3KS^4 \cos^2\beta \sin 2\theta (\sin^2\beta - \cos^2\beta \sin^2\theta)
\end{eqnarray}
which generally favors canted spin structures $0 < \theta < \pi /2$.

\section{Magnetic Phase Diagram of CuFeO$_2$} 

The nonlocal free energy functional formulated above can be expressed as a function of the parameters which characterize the spin density $F=F({\bf Q},S,m,\phi,\gamma,\theta,\beta)$.  Equilibrium spin configurations as a function of temperature and magnetic field are then determined by minimization (numerically) of $F$ for a given set of coefficients.  Since the phase diagram involves spin structures characterized by a number of different wave vectors ${\bf Q}$ which can each have associated umklapp terms, it is necessary to minimize (with respect to the other parameters) separately and compare the distinct free energies associated with IC, $3{\bf Q} = {\bf G}$, $4{\bf Q} = {\bf G}$, and $5{\bf Q} = {\bf G}$ (labeled $F_{IC}, F_{P3},F_{P4}, F_{P5}$) phases in order to determine the equilibrium state.  

A feature of the present model is the large number of coefficients which is a consequence of not only the nonlocal formulation but also the many competing interactions.  This multitude of effects are, however, essential for a complete understanding of the complex magnetic phase diagram of this highly frustrated compound. In order to reduce the number of free parameters, zero-T coefficient values from I are used here and other coefficients are {\it a priori} assigned reasonable values.  Only a relatively small number of coefficients are adjusted in an effort to reproduce the essential features of the phase diagram. As in I, the overall energy scale is set by assigning $J_1 \equiv 1$ with other exchange parameters given by $J_2=J_3=0.3$ and $J'=0.4$.  Magnetoelectric coupling coefficients used in I are adopted here,  $C_1=0.3$, $C'=0.1$ and $A_P=1$, as well as the assignment of an anisotropy strength of $3\%$ so that $J_{iz}=0.03J_i$ and $J'_z=0.03J'$. This level of anisotropy is also adopted here for the trigonal coefficient, $K=0.03$.  Some of the Landau coefficients were arbitrarily set as follows: $a=1$, $B_0=1$, $B_1=1$, $B_2=0.1$, $B_3=0.1$ and $C=0.1$.  Assuming a positive value for the term  $B_4 |{\bf m} \cdot {\bf S}|^2$ was always found to yield a low-field spin-flop transition (since anisotropy is weak), which is not observed in the magnetic phase diagram of CuFeO$_2$ (at least moderate field values). Assigning the negataive value $B_4=-0.2$ serves to enhance the stability of the reported configurations with ${\bf S} || {\bf H}$ in the P5 and P3 states (also see Ref.\onlinecite{plumer89}). (This point is discussed further below.) Coefficients of the umklapp terms, $B_{4,3}$, $B_{4,4}$, $C_{6,6} \equiv C_{6,3}$, $C_{6,4}$, and $C_{6,5}$ are then adjusted to reproduce essential features of the $H-T$ phase diagram.  Note that the $6{\bf Q}={\bf G}$ umklapp term in (31) contributes to the P3 state and its coefficient is assigned the same value as the sixth-order $3{\bf Q}={\bf G}$ umklapp term for simplicity.  

For the range of coefficients considered here, the inter-layer phase factor $\phi$ is always found to be zero.  The overall spin polarization phase factor $\gamma$ appears only in umklapp terms and the free energy is minimized with $\gamma = \pi /n$ for commensurate states described by $n{\bf Q} = {\bf G}$.

Consider first the sequence of transitions which occur at $H=0$. With the addition of axial anisotropy, the spin structure to first appear as the temperature is lowered from the paramagnetic state is linearly polarized ${\bf S} || {\bf \hat c}$.  The free energy up to fourth order is given by
\begin{eqnarray} 
F_0 = (A_{\bf Q} + J_{z{\bf Q}})S^2 + (B_1 + {\textstyle\frac{1}{2}} B_2)S^4 + {\textstyle\frac{1}{2}}B_{4,4}S^4 \cos(4\gamma)\Delta_{4{\bf Q},{\bf G}}
\end{eqnarray}
with the transition temperature given by $T_{N1} = - (J_{\bf Q} + J_{z{\bf Q}})/a$.  The wave vector is thus determined by the value which maximizes this function, as shown in Fig. 1.  For the set of exchange parameters used here, this gives the collinear IC phase with degenerate wave vectors $(q_x,q_y,q_z) = (2q,2 \pi,\pi)$, $(2\pi-q,3/2 q-\pi,\pi)$ and $(2\pi-2q,\pi,\pi)$ (or equivalently $(q_a,q_b,q_c)= (2q, 2 \pi +q,\pi), (2\pi-q,q,\pi)$, and $(2\pi-2q,2\pi-q,\pi)$ respectively) where $q \simeq 0.22$, giving $T_{N1} \simeq 2.0$.  The result $\frac{1}{5} < q < \frac{1}{4}$ is consistent with neutron diffraction data\cite{mitsuda00} and the speculation of multi-${\bf q}$ domain structures.\cite{mitamura07} Numerical minimization of the free energies corresponding to IC and P4 $(\pi, 2\pi, \pi)$ phases with all terms included is then performed in order to compare $F_{IC}$ and $F_{P4}$ as a function of temperature.\cite{sato}  Assigning the parameters values $B_{4,4} = 0.8$ and $C_{6,4}=0.2$ is found to yield the result that $F_{P4}< F_{IC}$ for $T < T_{N2} \simeq 1.5$, which approximately agrees with the experimental data for $T_{N2} / T_{N1}$. The transition at $T_{N1}$ is continuous while the IC-P4 transition at $T_{N2}$ is discontinuous.

Using this method of comparing free energies associated with IC and commensurate spin states, the phase diagram with ${\bf H} || {\bf \hat c}$ is determined numerically.  The three remaining coefficients are fit to best reproduce the transition boundaries in an effort to achieve semi-quantitative agreement with the corresponding experimental results presented in Refs.\onlinecite{kimura06} and \onlinecite{terada06}. This procedure yields $B_{4,3} = 1.0$,  $C_{6,6} =0.1$, and $C_{6,5}=0.6$.  The linear IC, P5 and P3 phases remains stable with ${\bf S} || {\bf \hat c}$ at all field values considered due to setting $B_4 < 0$.  The elliptical IC phase is stabilized at moderate field values with $\beta \simeq 0.19 \pi$ at lower T.  There is little change in wave vector from the zero field values since $C_x^2$ and $C_y^2$ in (36) are small.  Due to the small trigonal interaction term $F_K$, this structure is also found to be canted away from the ${\bf \hat c}$-axis by about $\theta \simeq 10^0$.  Similarly, at low temperatures, the linear P4 phase exhibits a discontinuous transition to canting away from the ${\bf \hat c}$ axis by about $45^0$ (indicated by the thin solid line in Fig. 2).  The transition between elliptical and linear IC phases is continuous with $\beta$ acting as an order parameter.  The canting of magnetic structures does not occur within the present model in the case $K=0$.  Critical fields at low temperature $H_{P4-IC} \simeq 1$, $H_{IC-P5} \simeq 2$, $H_{P5-P3} \simeq 3$ are in fair agreement with the experimentally observed ratios  $H_{IC-P5} / H_{P4-IC} \simeq 13T/7T$ and $H_{P5-P3} / H_{P4-IC} \simeq 20T/7T$.\cite{kimura06,terada06}

 \begin{figure}[h]
\includegraphics[width=1.0\textwidth]{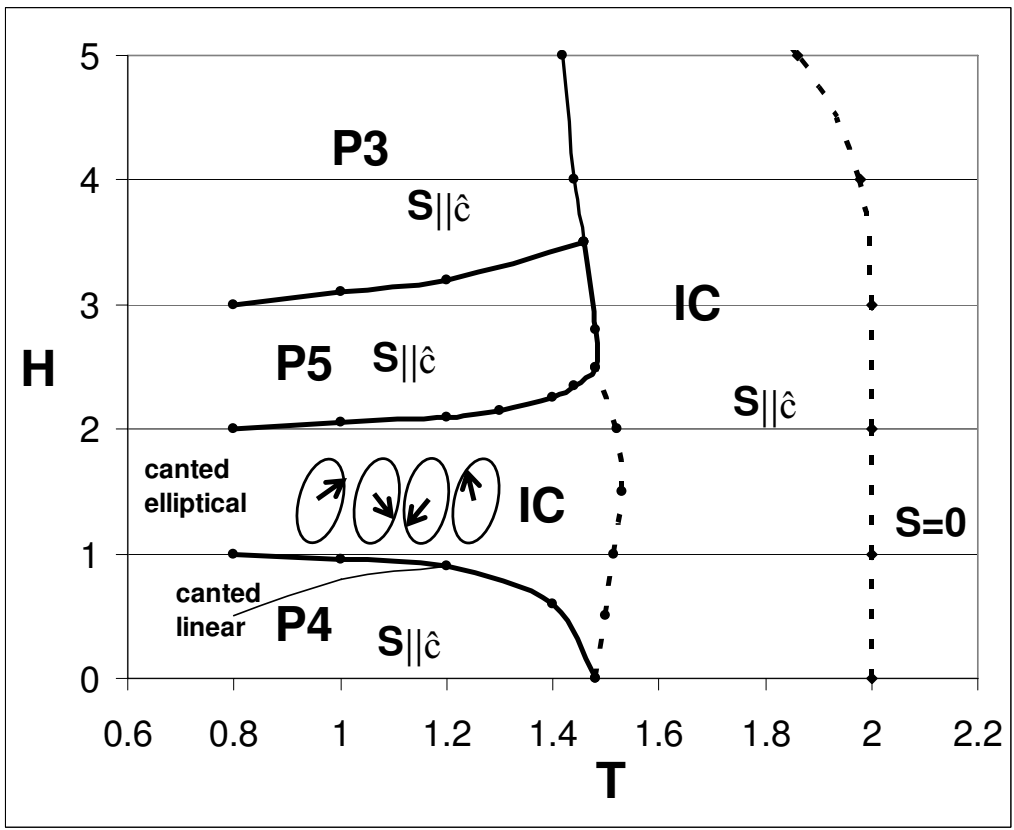}
\caption{\label{one} Sketch of the model phase diagram based for CuFeO$_2$ on minimization of the free energy (in units of $J_1$). Solid and broken lines represent discontinuous and continuous transitions, respectively.}
\end{figure}

\section{Discussion and Conclusions} 
 
In this work, a model nonlocal Landau-type free energy has been developed and analyzed that captures essential features of the complex magnetic phase diagram of the highly frustrated magnetoelectric triangular antiferromagnet CuFeO$_2$ and serves as an extension of a previous zero temperature model.\cite{plumer07}  In contrast with local formulations, the present approach naturally leads to the result that each symmetry-invariant isotropic term has its own independent coefficient.  In the case of CuFeO$_2$, it is argued that strong magnetoelastic coupling is one source for such nonlocal biquadratic (and higher order) spin interactions.  As emphasized in Ref. \onlinecite{plumer07}, magnetoelectric coupling also provides a microscopic mechanism for the existence of biquadratic antisymmetric exchange. In addition to weak axial anisotropy, the rhombohedral crystal symmetry also supports a somewhat unusual trigonal term which couples basal-plane and axial spin components.  With analysis performed in terms of a Fourier expansion of the spin density, this approach also facilitates comparison with neutron diffraction data.  

There are a number of limitations associated with this method. The Landau model is an expansion of the free energy around ${\bf s(r)} = 0$ so that results far below $T_N$ and at high field values may require a large number of terms.  In addition, only a small number of modulated spin structures can be fully described by just a few Fourier components.  The model in its present form fails, for example, to account for the very high field spin-flop P3 phase suggested by magnetization measurements.\cite{terada06} The merging of P4, IC and P5 transition boundaries suggested by the experimentally determined phase diagram is also not found in our truncated model. In addition, it is a mean field theory.  Within the present formulation, the paramagnetic-IC (para-IC)  transition temperature is given by $T_{N1} = -(J_{\bf Q}+ J_{z{\bf Q}})j^2/(ak_B)$ where $a=3/[2j(j+1)]$.  Using $j=5/2$ and the estimate $J_1 \simeq 5.3K(k_B)$ from spin-wave data,\cite{ye07} (along with the other exchange interactions values given above) gives $T_{N1} = 38.6 K$.  The nearly factor of 3 discrepancy with the experimental value can be attributed to strong critical fluctuation effects associated with the high degree of frustration in this compound.   

The nature of the symmetry breaking at the two continuous transitions, para-IC and IC(linear)-IC(elliptical), can be analyzed in a manner similar to the weak axial triangular antiferromagnet CsNiCl$_3$.\cite{kawa}  Both should belong to the standard $XY$ universality class involving a two component continuous order parameter.  In the case of the para-IC transition, the two components can be identified as the magnitude of the spin order, $S$, and the associated phase angle $\gamma$.  In the case of unfrustrated systems, this transition would be identified with Ising universality.  In the finite-field transition to the elliptical phase, the system develops (XY) basal-plane components ${\hat {\bm \rho}_1}$ and ${\hat {\bm \rho}_2}$.

A consequence of the present model and analysis is that basal-pane components of the electric polarization ${\bf P}$ are induced by the mechanism proposed by Kimura {\it et al.}\cite{kimura06} if the spin structure is canted relative to the ${\bf \hat c}$ axis.  This result offers a resolution to the issue raised recently concerning the microscopic origin of the magnetolectric effect in this compound.\cite{nakajima07,terada07,arima07,mitamura07}  Such a canting occurs in crystals with rhomboderal symmetry only as a consequence of the trigonal interaction term $F_K$.

The method, and some of the results presented here, are also relevant to other frustrated triangular antiferromagnets.  Although  hexagonal MnBr$_2$ shows planar anisotropy and strong inter-plane coupling responsible for a period-4 $c$-axis modulation, it shares the same basal-plane P4-IC transition as CuFeO$_2$.\cite{sato} NaFeO$_2$ has the same rhombohedral symmetry as CuFeO$_2$ and a P4-IC transition but also with ${\bf S} \perp {\bf \hat c}$ and a more complicated P4 spin structure.\cite{mcqueen} Two of the three basal-plane triangular crystal directions show period-4 modulations and the third is ferromagnetic.  This can be described by ${\bf q} = (\pi/2,\pi/4)_{xy}$ or equivalently ${\bf q} = (\pi/2,\pi/2)_{ab}$, which satisfy the umklapp requirement $4{\bf Q}  = {\bf G}$.  Differences with CuFeO$_2$ could be due to additional inter-plane exchange interactions. It is of interest to note that this material does not appear to exhibit the structural phase transition to monoclinic symmetry found in CuFeO$_2$ below $T_{N1}$, providing another example where the stability of a P4 phase is not related to a structural distortion.   Other potential applications of the present model include the hexagonal magnetoelectric antiferromagnet RbFe(MoO$_4$)$_2$ which exhibits a complex phase diagram involving  P3, P4 and P5 spin structures,\cite{rbfe} and a series of rhombohedral magnetoelectric antiferromagnets $A$CrO$_2$ (A=C$_r$, A$_g$, Li, or Na).\cite{acro}

\section{Acknowledgments}
I thank  G. Quirion and M.J. Tagore for enlightening discussions. This work was supported by the Natural
Science and Engineering Research Council of Canada (NSERC).

\end{document}